\documentclass[twocolumn,aps,prb,superscriptaddress]{revtex4-1}                                                          

\usepackage{amsmath,amssymb,amsthm}                                                                                
\usepackage{mathtools}                                                                                             
\usepackage{xspace}                                                                                                
\usepackage[usenames,dvipsnames]{xcolor}                                                                           
\usepackage[colorlinks,hyperindex,breaklinks]{hyperref}                                                            
\usepackage{braket}                                                                                                
\usepackage{hyphenat}                                                                                              
\usepackage{bbold}                                                                                                 
\usepackage{enumerate}                                                                                             
\usepackage{tabu,booktabs}                                                                                         
\usepackage{tabularx}                                                                                              
\usepackage[verbose]{placeins}                                                                                     
\usepackage[hang,flushmargin]{footmisc}                                                                            
\usepackage{comment}                                                                                               
                                                                                                                   
\usepackage[ngerman,british]{babel}

\usepackage{srcltx}                                                                                                

\renewcommand{\arraystretch}{1.6}

\renewcommand{\vec}[1]{\mathbf{#1}}              

\begin{document}

\title{Impact of finite temperatures on the transport properties of Gd\\ from first principles}

\author{K. Chadova}
\email{kchpc@cup.uni-muenchen.de}                                                                                  
\affiliation{Department Chemie, Ludwig-Maximilian-University Munich, Butenandtstrasse 5-13, 81377 Munich, Germany}
\author{S. Mankovsky}                                                                               
\affiliation{Department Chemie, Ludwig-Maximilian-University Munich, Butenandtstrasse 5-13, 81377 Munich, Germany}
\author{J. Min\'ar} 
\affiliation{Department Chemie, Ludwig-Maximilian-University Munich, Butenandtstrasse 5-13, 81377 Munich, Germany}
\affiliation{New Technologies-Research Center, University of West Bohemia, Univerzitni 8, 306 14 Pilsen, Czech 
Republic}        
\author{H. Ebert}                                                                               
\affiliation{Department Chemie, Ludwig-Maximilian-University Munich, Butenandtstrasse 5-13, 81377 Munich, Germany}
\date{\today}


\begin{abstract}
Finite temperature effects have a pronounced impact on the transport properties of solids. In magnetic systems, 
besides the scattering of conduction electrons by impurities and phonons, an additional scattering source coming from 
the magnetic degrees of freedom must be taken into account. A first-principle scheme which treats all these scattering
effects on equal footing was recently suggested within the framework of the multiple scattering formalism. Employing 
the alloy analogy model treated by means of the CPA, thermal lattice vibrations and spin fluctuations are 
effectively taken into account. In the present work the temperature dependence of the longitudinal resistivity and the 
anomalous Hall effect in the strongly correlated metal Gd is considered. The comparison with experiments demonstrates 
that the proposed numerical scheme does provide an adequate description of the electronic transport at finite 
temperatures.
\end{abstract}


\maketitle

\renewcommand{\arraystretch}{1.6}
\newcommand{\dd}{\mathrm d}
\newcommand{\del}{\partial}

\newcommand{\matr}[1]{\mathbf{\underline{#1}}}
\newcommand{\matrg}[1]{\underline{\boldsymbol{#1}}}

\newcommand{\ee}{\text e}
\newcommand{\ii}{\text i}

\renewcommand{\vec}[1]{\mathbf{#1}}
\newcommand{\spacegroup}[1]{$#1$}

\newcommand{\tableheading}[1]{\textbf{#1}}

\newcommand{\tensordir}{../sym/tmp/}
\newcommand{\matrixtablelinesep}{0.55mm}
\newcommand{\matrixtablearraystretch}{0.9}
\newcommand{\matrixtablearraycolsep}{0.3pt}

\newcommand{\TR}{\mbox{Tr}}
\newcommand{\SIGTEN}{\mbox{$\matrg \sigma$}}
\newcommand{\TORTEN}{\mbox{$\matrg t$}}

\section{Introduction \label{sec:I}}
Rare earth elements may exhibit both ferromagnetic or antiferromagnetic order in certain temperature regimes. Nowadays, 
it is commonly accepted that Gd, having the hcp structure, possesses a simple ferromagnetic (FM) order up to its 
Curie temperature ($T_c$). However, in early experimental studies a helical magnetic structure was observed in 
polycrystalline Gd in the temperature range between $210$ K and $290$ K~\cite{BP62}. Such a helical spin configuration 
is easily destroyed by a weak magnetic field~\cite{BP62} leading to a collinear magnetic structure in the system. This 
means that only in the absence of an applied magnetic fields this type of antiferromagnetism can be observed. Recent 
experiments on single crystals of Gd did not reveal any anomalies in the low-field magnetization curves and confirm that 
Gd has a normal ferromagnetic structure up to its Curie temperature~\cite{KS00,WNA64}. The Curie temperature 
determined experimentally was found to be $289$ K with a saturated magnetic moment of $7.12\mu_B$~\cite{Trombe37,ELS53}. 
In another experimental study the Curie temperature was determined to be $293.2$ K  with an absolute saturation moment 
of $7.55\mu_B$~\cite{NLS63}. Although Gd behaves like a simple ferromagnet it has nevertheless a rather complex 
temperature dependence of its magnetization: as the temperature decreases to $230$ a spin-reorientation occurs from the 
magnetization parallel to the $c$ axis to the magnetization tilted by $30^{\circ}$ with respect to the $c$ axis, reaching 
its maximum tilt angle of $60^{\circ}$ at around $T=180 K$~\cite{KS00}. Such a behavior is quite demanding concerning an 
adequate theoretical description. Therefore, in the present work the direction of the magnetization is taken along the 
$c$ axis unless it is mentioned otherwise.

It is well established that the magnetism in Gd is dominated by $f$-electrons with a magnetic moment of $7 \mu_B$ due 
to half-filling of the highly localized $4f$ states. The observed excessive magnetic moment is attributed to the 
valence $5d6sp$ band exhibiting spin polarization due to the strong exchange field created by the $4f$-electrons~\cite{RCM+75} 
as it is extensively discussed in the literature~\cite{BAE+92,WSM+96,MMD+02,GBK+12a,ODS+15}. In particular, these 
discussions concern the finite temperature behavior of the magnetic moment of the valence electrons~\cite{KTM04} 
observed experimentally. In earlier discussions it has been suggested to treat these on the basis of the Stoner 
model~\cite{DGP96}. Recent investigations by experiment~\cite{MMD+02,LPB+95,FVL+02} as well as 
theory~\cite{GBK+12a,San14,ODS+15} based on first-principles calculations clearly demonstrate the finite exchange 
spitting of valence states above the Curie temperature despite the vanishing total magnetization, which implies a much 
more complicated picture of interactions than provided by the simple Stoner model. 

The rather different origin of the spin magnetic moment for the $f$- and $5d6sp$-electrons leads also to a 
different dynamical behavior characterized in general by a different magnetization dissipation rate. 
This would imply separate spin dynamics equations for $f$- and $5d6sp$-spin magnetic moments coupled via the exchange 
interactions, as was considered in particular in Gilbert damping calculations by Seib and F\"ahnle~\cite{SF10}. The 
authors, however, point out that the common equation for all types of spin moments can be used in the limit of slow 
magnetization dynamics~\cite{SSF09}, that allows to use also a common Gilbert damping parameter calculated within 
the adiabatic approximation.    

It is well-known, that in magnetic systems the electrical resistivity is caused by electron scattering by various 
magnetic inhomogeneities in addition to the electron-phonon scattering as well as scattering by impurities and other 
structural defects. The latter contribution is responsible for the so-called residual resistivity observed in the 
zero-temperature limit. The resistivity part due to the phonon mechanism shows usually a $~T^5$ behavior at low 
temperatures and varies linearly with $T$ above the Debye temperature ($T_D$). This behavior can be described on the 
\textit{ab initio} level and corresponding studies on transition metals~\cite{SS96} lead in general to good agreement 
with experimental data. In the present study not only the linear dependence was obtained in the temperature region $T > 
T_D$ but it was found also well below $T_D$. A theoretical description of the resistivity caused by thermal 
spin-fluctuation effects was first given on the basis of the $s$-$d$ (in rare earth $d$-$f$) model 
Hamiltonian~\cite{Kas56,Kas59,GF58}. This approach suggests a $T^2$ dependence in the low temperature limit and almost 
a constant resistivity above the Curie temperature. In the intermediate temperature regime the $T$-dependence of the 
resistivity is expected to be rather complex. Recent {\it ab initio} calculations of the paramagnetic spin-disorder 
resistivity for a number of transition metals and their alloys as well as rare earth metals are based on two alternative 
approaches: the disordered local moment approach using the coherent potential approximation (CPA) formalism and 
averaging the Landauer-B\"uttiker conductance of a supercell over the random non-collinear spin-disorder configurations, 
with both leading in general to good agreement with experimental values~\cite{KDT+12,GBK+12a}. However, for a quantitative 
description of the temperature dependent electrical resistivity from first principles one needs to combine 
the influence of lattice vibrations and spin fluctuations which is a non-trivial task. Therefore, certain approximations 
are required to reach this goal.

During the last years, the anomalous Hall effect (AHE) and its dependence on the temperature attracts also much 
attention. In the case of Gd, a number of theoretical investigations have been performed to explain the unexpectedly 
large AHE observed experimentally \cite{CGS79}. Previously, these studies were performed on a model level. An earlier 
description of the AHE of Gd was based on the uniform electron gas model accounting for spin-orbit coupling effects 
leading, in turn, to an asymmetry in the scattering process (skew-scattering mechanism)~\cite{KL54}. However, due to 
the high localization the electrons giving rise to the magnetic moment are unable to participate in conduction, 
therefore this model is not appropriate to describe the AHE in rare earth systems. The model developed by 
Kondo~\cite{Kon62} was based on the $s$-$d$ ($s$-$f$) interaction leading to a scattering of the conduction electrons by 
the thermally induced spin moment tilting. In this model the necessary asymmetry is due to the intrinsic spin-orbit 
coupling of the $f$ electrons. Therefore, the Hamiltonian describing the interaction of the conduction and the 
localized electrons is valid when the orbital angular moment of localized electrons remains unquenched. This is not the 
case for Gd and therefore it cannot be used to describe the AHE in this metal. Another model which eliminated the above 
mentioned constraint, was developed by Maranzana~\cite{Mar67} and is based on Kondo's model. In this model the 
skew-scattering mechanism is originating from the interaction between the localized spin moment and the orbital momentum 
of the conduction electron.

Within the discussed models the large AHE in Gd was ascribed solely to the skew-scattering contribution. Another 
scattering mechanism, the so-called side-jump mechanism, first introduced by Berger~\cite{Ber70, Ber72} was accounted 
within a model suggested by Fert~\cite{AFR83}. It was demonstrated, particularly for Gd, that the side-jump contribution 
is equally important as the skew-scattering mechanism and should be taken into account as well.

In this paper, we discuss the impact of finite temperatures, taking into account thermal lattice vibrations and spin 
fluctuations, on the transport properties in Gd from first principles by making use of the alloy analogy 
model~\cite{EMC+15}.

\section{Computational details \label{sec:F}}

The electronic structure calculations are based on KKR Green function method~\cite{EKM11} implemented in the fully 
relativistic spin-polarized Munich SPR-KKR package with angular momentum cutoff $l_{max} =4$. A full four-component 
Dirac formalism is employed to describe the electronic structure within Kohn-Sham-Dirac density functional 
theory~\cite{ED11}. For $spd$ electrons the local density approximation was used with the parametrization given by 
Vosko {\em et al.}~\cite{VWN80}. To treat the highly correlated 4$f$ states the LSDA+$U$ method was used with the double 
counting part of the LSDA+$U$ functional evaluated within the so-called atomic limit expression~\cite{CS94}. The 
temperature effects are treated within the alloy analogy scheme~\cite{EMC+15} based on the CPA alloy 
theory~\cite{Vel69,But85,TKD+02b}. For the description of the magnetic spin fluctuations the temperature-dependent 
magnetization data was taken from experiment~\cite{NLS63}. 
The calculation of the transport properties of Gd is based on the Kubo-St\v{r}eda formalism, with the corresponding 
expression for the conductivity given by: 
%
\begin{eqnarray}                                                                                                   
\label{eq:Kubo-streda}                                                                                             
\sigma_{\rm \mu\nu}                                                                                                
&                                                                                                                  
=                                                                                                                  
&                                                                                                                  
\frac{\hbar }{4\pi N\Omega}                                                                                        
{\rm Trace}\,\big\langle \hat{j}_{\rm \mu} (G^+-G^-)                                                               
\hat{j}_{\rm \nu}  G^-                                                                                             
\nonumber                                                                                                          
\\                                                                                                                 
&&                                                                                                                 
\qquad \qquad \quad                                                                                                
-  \hat{j}_{\rm \mu} G^+\hat{j}_{\rm \nu}(G^+-G^-)\big\rangle_{\rm c}                                              
\nonumber                                                                                                          
\\                                                                                                                 
&&                                                                                                                 
+ \frac{|e|}{4\pi i N\Omega} {\rm Trace}\,                                                                         
\big\langle (G^+-G^-)(\hat{r}_{\rm \mu}\hat{j}_{\rm \nu}                                                           
- \hat{r}_{\rm \nu}\hat{j}_{\rm \mu}) \big\rangle_{\rm c}                                                          
\label{eq:bru}                                                                                                     
\;                                                                                                               
\end{eqnarray} 
with the relativistic current operator $\hat{\vec  j} =  -  |e| c \vec{\mbox{\boldmath{$\alpha$}}}$ and the 
electronic retarded and advanced Green functions $G^{\pm}$, evaluated at the Fermi energy $E_F$ by means of the 
relativistic multiple scattering or KKR formalism~\cite{EKM11}. The angular brackets denote a configurational average 
which here is carried out using the coherent potential approximation (CPA) which takes into account the so-called 
vertex corrections (VC)~\cite{But85}. In the last equation $N$ is the number of sites and $\Omega$ is the volume of the 
unit cell. As was justified by previous work~\cite{NHK10} the second term in the Eq.(\ref{eq:bru}) has been omitted.

The Gilbert damping parameter~\cite{BTB08,EMKK11} was calculated within the linear response theory using the  
Kubo-Greenwood-like equation:
%
\begin{equation}
\label{eq:sig}
\alpha_{\mu\nu}  
= -\frac{\hbar\gamma}{\pi M_s} 
{\rm{Tr}}\,
\big\langle \hat{T}_{\mu} \, \Im G^+\, \hat{T}_{\nu} \, \Im G^+ \big\rangle_{\rm c} \;,
\end{equation}
%
where $M_s$ is the saturation magnetization, $\gamma$ the gyromagnetic ratio and $\hat{T}_{\mu}$ is the torque 
operator~\cite{EMKK11}.

\begin{figure}[t]      
 \begin{center}
  \includegraphics[angle=0,width=1\linewidth,clip]{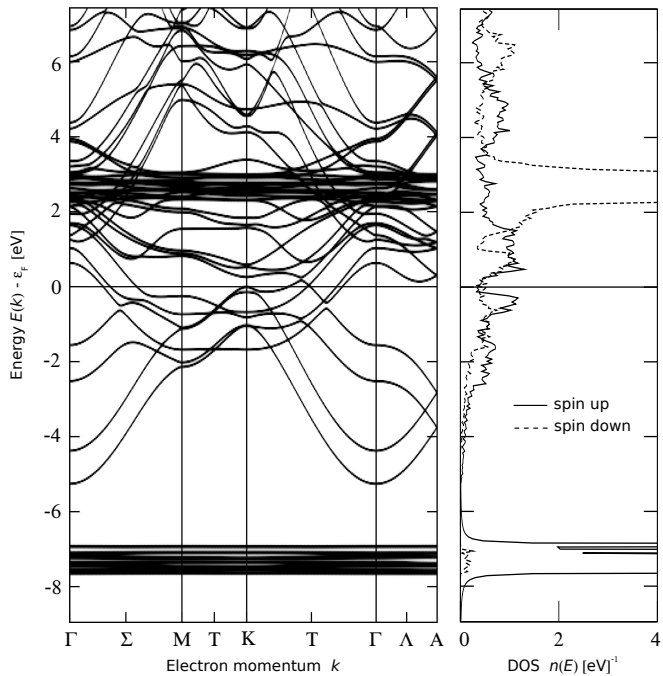}
    \caption{\label{plot:Gd_bsf_dos} (Color online) Band structure and the density of states in Gd, calculated using 
LSDA+$U$ approach.}
\end{center}
\end{figure}

\section{Results \label{sec:R}}

\subsection{Electronic structure}

The electronic structure of Gd has been calculated using the experimental lattice parameters $a = 3.629$\AA, $c/a=1.597$. As 
was mentioned above, the $4f$ electrons have been treated as the valence electrons with correlations 
described within the LSDA+$U$ scheme with the Coulomb parameter $U$ = 6 eV and the exchange parameter $J$=0.9 eV. The 
resulting electronic band structure is shown in Fig.~\ref{plot:Gd_bsf_dos} in combination with the density of states (DOS). 
One can see that the $4f$ majority-spin states are located at approximately $-7.5$ eV with respect to the Fermi level ($E_F$), 
while the minority-spin states are at about $3$ eV above $E_F$, which is in agreement with photoemission 
experiments~\cite{LBC81}. 

\begin{figure}[h!]
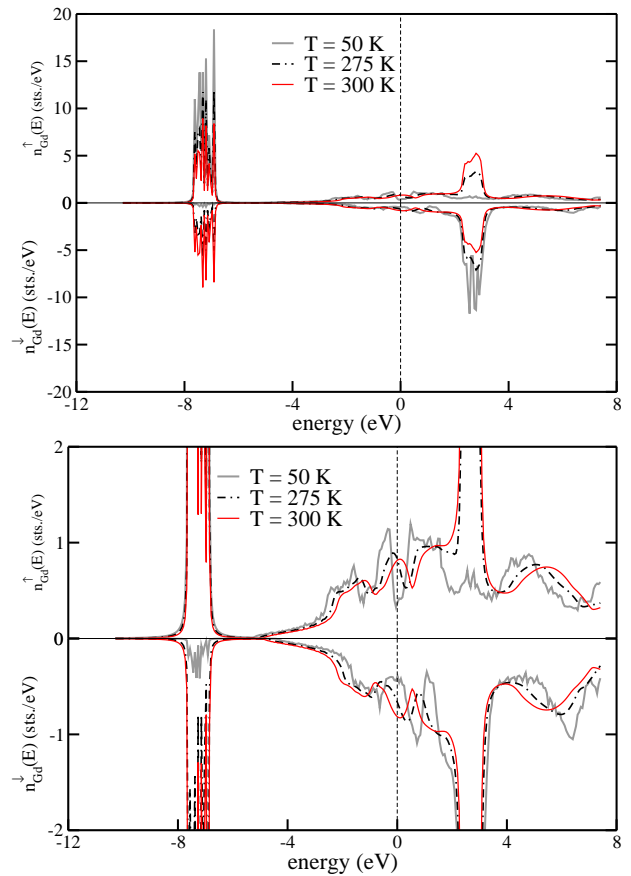

\begin{center}
 \includegraphics[angle=0,width=0.94\linewidth,clip]{Gd_DOS_tot_T.eps}\\
 \includegraphics[angle=0,width=0.94\linewidth,clip]{Gd_DOS_tot_zoom_T.eps}
   \caption{\label{fig:DOS} (Color online)  Spin resolved DOS of Gd for various temperatures: bottom panel 
- magnified area.}
\end{center}
\end{figure}

The spin magnetic moment obtained in the calculations for $T = 0$~K equals to $7.63 \mu_{B}$ and accordingly is in a 
good agreement with the experimental saturated magnetic moment of $7.55 \mu_{B}$/per atom~\cite{NLS63}. The dominating 
contribution of $7 \mu_B$ is associated with the $f$ electrons, while the excessive spin magnetic moment of $0.63 
\mu_{B}$ is a result of the exchange splitting for the $5d6s6p$ electrons due to a strong exchange field produced by the 
$f$-electrons, as it was discussed previously~\cite{RCM+75,KTM04,ODS+15}. The persistence or vanishing of the exchange 
splitting with increasing temperature is a matter of debate both in theory and experiment. Several experimental reports 
indicate that it collapses approaching the Curie temperature~\cite{KAE+92}, while other demonstrate that the exchange 
splitting persists even in the paramagnetic state~\cite{LZD+92,MMD+02}. The spin resolved 
total DOS calculated in the global frame of reference with the quantisation axis along the average magnetization at 
finite temperatures is represented in Fig.~\ref{fig:DOS}. Obviously, a temperature increase results in changes of 
the majority and minority spin DOS due to the spin mixing caused by the 
thermal spin fluctuations. This leads to the same DOS for both spin directions at $T>T_c$. The energy positions of the 
$f$-states are almost unchanged in the whole temperature region. However, the exchange splitting of the spin-up and 
spin-down $5d6s6p$-states (having the main contribution to the DOS at the energies around $E_F$) decreases (as it 
depends on the average magnetization of the system) with increasing temperature. In particular, this results in an increase 
of the DOS at the Fermi level in the paramagnetic state.

\subsection{Electrical resistivity}

\begin{figure}
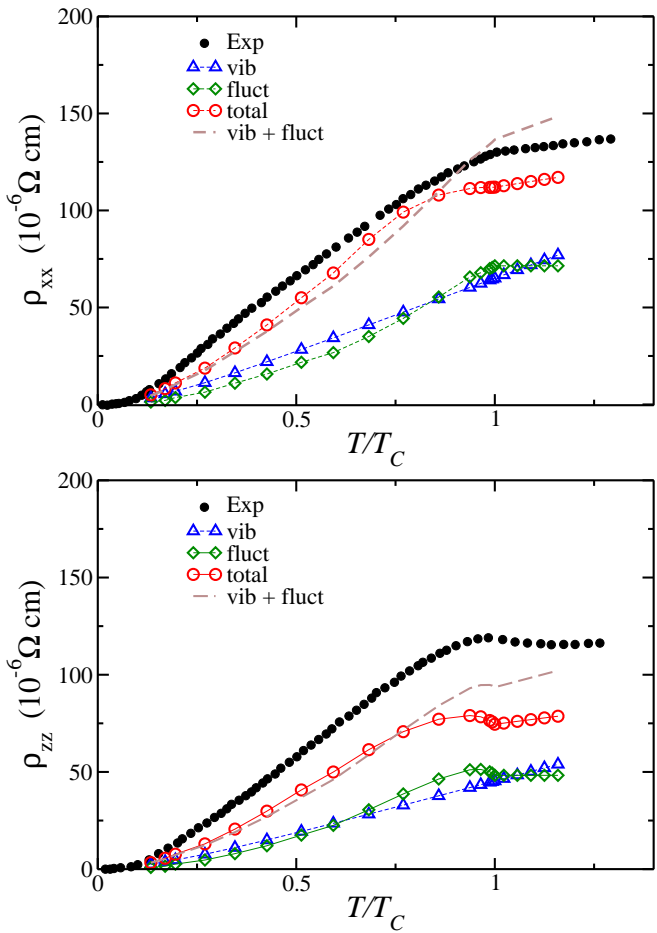

 \begin{center}
 \includegraphics[angle=0,width=1\linewidth,clip]{Gd_RHO_xx_contrib_LDAU.eps}\; \\
 \includegraphics[angle=0,width=1\linewidth,clip]{Gd_RHO_zz_contrib_LDAU.eps}\;
   \caption{\label{plot:Gd_rho_xx_zz} (Color online) Temperature-dependent electrical resistivity: top panel in-plane, 
bottom panel - out of plane components. The various symbols represent: black solid circles -- experimental 
results~\cite{NLS63}, empty blue triangles -- only thermal lattice vibrations, empty green diamonds -- only spin 
fluctuations, empty red circles -- total resistivity including both effects simultaneously, brown dashed line 
corresponds to the sum of individual contributions.}
\end{center}
\end{figure}

One of the central transport properties of metallic systems is their electrical resistivity. The experimentally 
measured temperature-dependent resistivity of Gd exhibits an anisotropy with different magnitudes along the hexagonal 
axis ($\rho_{zz}$) and in the basal plane ($\rho_{xx}$)~\cite{NLS63} (see Fig.~\ref{plot:Gd_rho_xx_zz}). Both $\rho(T)$ 
curves are characterized by an abrupt slope change close to the Curie temperature.
%

In addition to the total $\rho(T)$ values, we have investigated its temperature dependence caused only by lattice 
vibrations (vib) or only by magnetic fluctuations (fluct), which appear to be of comparable magnitude. From this one 
has to conclude that these sources of the temperature-dependent resistivity are additive only in the case of 
the weak disorder (low temperatures), which does not hold when approaching the Curie temperature (strong disorder). 
In this regime they must be taken into account simultaneously, since only then the overall behavior of the resistivity 
curves agrees well with experiment. This allows to conclude that the maximum of the experimental $\rho_{zz}$ (close to 
the Curie temperature) is not a result of short range magnetic order as it was suggested in the earlier 
literature~\cite{NLS63}, since the present calculations are based on the single-site CPA.
\begin{figure}
 \begin{center}
 \includegraphics[angle=0,width=1\linewidth,clip]{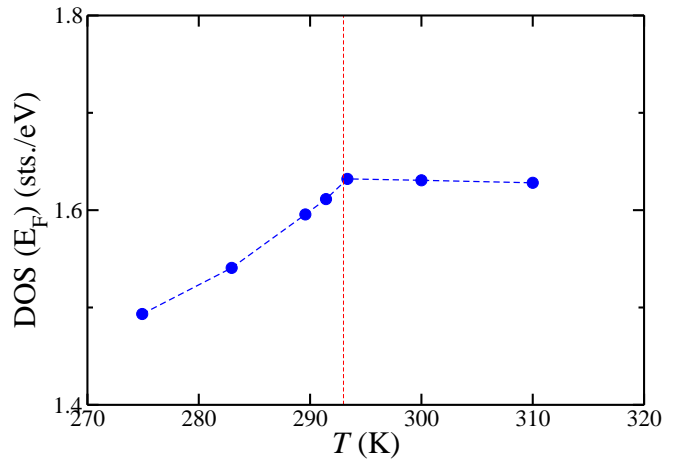} 
   \caption{\label{plot:Gd_DOS_tot_EF} (Color online) Total DOS at the Fermi level depending on the temperature.}
\end{center}
\end{figure}
The present results suggest its origin as a combination of two competitive mechanisms. On the one hand, thermally 
induced disorder grows, leading to a resistivity increase and on the other hand, the effective DOS around $E_F$ 
relevant for the conductivity increases with increasing $T<T_c$ (Fig.~\ref{plot:Gd_DOS_tot_EF}), which 
effectively reduces the resistivity. 


\subsection{Anomalous Hall effect}

\begin{figure}
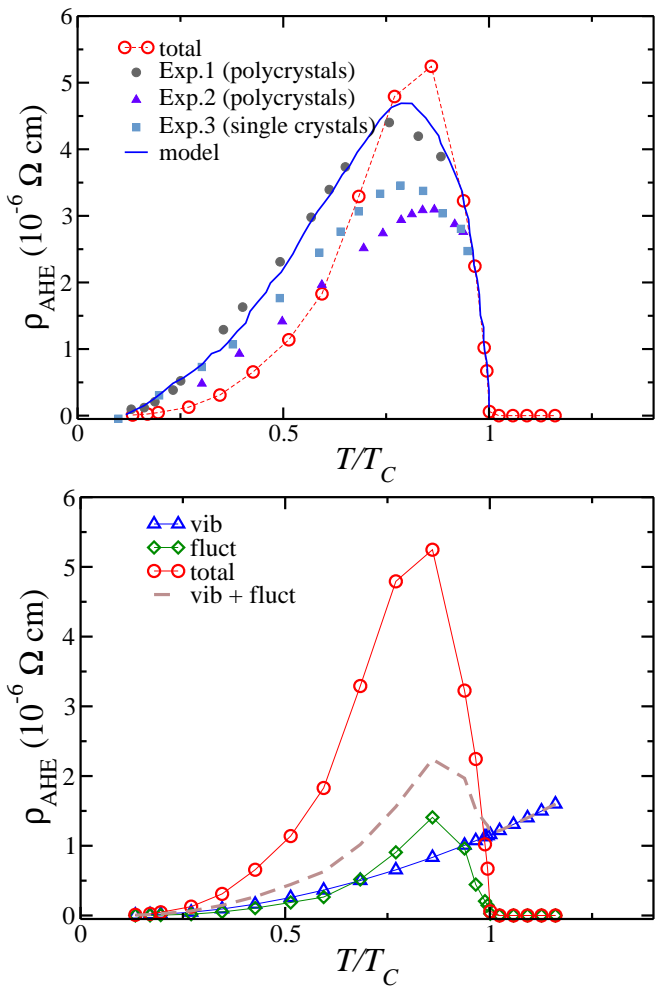

 \begin{center}
 \includegraphics[angle=0,width=1\linewidth,clip]{Gd_AHE_reduced_T_exp.eps}
 \includegraphics[angle=0,width=1\linewidth,clip]{Gd_AHE_reduced_T_KKR.eps}
   \caption{\label{plot:AHE} (Color online) Anomalous Hall resistivity depending on the temperature: top panel in 
comparison with experimental results (Exp.1 -- Ref.~\cite{Bab66}, Exp.2 -- Ref.~\cite{Bab66}, Exp.3 -- 
Ref.~\cite{LL67,VGF70}) and results from model calculations~\cite{AFR83}, bottom panel - individual contributions. The 
used symbols represent: empty blue triangles -- only thermal lattice vibrations, empty green diamonds -- only spin 
fluctuations, empty red circles -- total resistivity including both effects simultaneously, brown dashed line 
corresponds to the sum of individual contributions.}
\end{center}
\end{figure}

As was already mentioned, Gd shows a rather large AHE, which is well described within a model that accounts 
at the same time for skew-scattering and side jump mechanisms~\cite{AFR83}. However, within this model only the 
electron scattering by thermally induced spin fluctuations is discussed, while the contribution from the 
electron-phonon mechanism is completely neglected. Within the present calculations both contributions are taken into 
account. The resulting total anomalous Hall resistivity can be seen in Fig.~\ref{plot:AHE} (top panel) in comparison 
with experimental results (for polycrystalline samples as well as single crystals) and the theoretical result obtained 
on the basis of model calculations by Fert~\cite{AFR83}. One can see that the anomalous Hall resistivity shows a 
pronounced temperature dependence: the resistivity increases from zero at $T=0K$ to a maximum value just below the 
Curie temperature and then drops to zero as the magnetization vanishes with further increasing temperature. Overall 
there is a qualitative and quantitative agreement of our first principles results with experiment as well as with 
the model calculations. In Fig.~\ref{plot:AHE} (bottom panel) the individual contributions arising from the scattering 
by the lattice vibrations and spin fluctuations are shown. One can see that both mechanisms provide contributions nearly 
of the same order of magnitude. The qualitative behavior of the total AHR is determined by the scattering due to spin 
disorder, while the contribution due to lattice vibrations shows, as expected, a monotonous increase with temperature. 
It is interesting to compare the sum of the individual contributions with the total AHR. From Fig.~\ref{plot:AHE} 
(bottom panel) one can see that the total AHR significantly exceeds the sum of these contributions. Therefore for the 
correct description of the total AHR it is necessary to account simultaneously for the combination of scattering due to 
the thermal lattice vibrations and spin fluctuations.

\subsection{Gilbert damping}

\begin{figure}[t!]
 \begin{center}
 \includegraphics[angle=0,width=1\linewidth,clip]{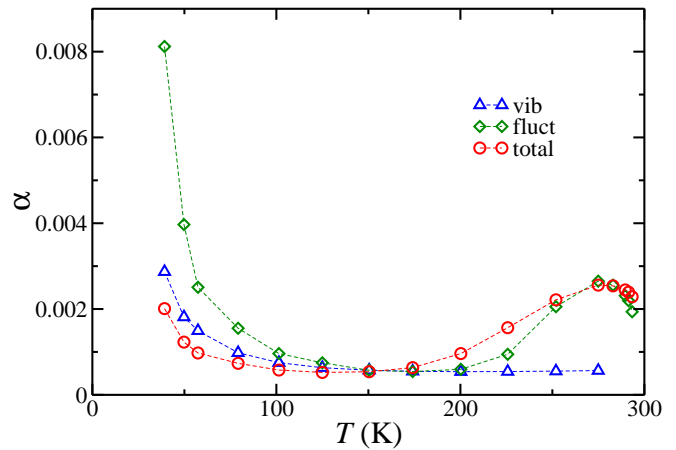}
   \caption{\label{plot:GD} (Color online) Gilbert damping parameter in Gd represented as a function of 
temperature. The used symbols represent: empty blue triangles -- only thermal lattice vibrations, empty green diamonds 
-- only spin fluctuations, empty red circles -- total resistivity including both effects simultaneously.}
\end{center}
\end{figure}

In the present work, the Gilbert damping parameter for Gd has been calculated in the limit of slow magnetization 
dynamics~\cite{SSF09}. It describes the magnetization dissipation for the whole system, accounting for $f$-like and 
$5d6sp$-like spin magnetic moments characterized by their slow simultaneous coherent motion. The corresponding results 
of calculations of the Gilbert damping as a function of temperature up to the Curie temperature are shown in 
Fig.~\ref{plot:GD}. The separate contributions due to thermal lattice vibrations and spin fluctuations are shown 
together with the curve accounting for both sources simultaneously. One can see a monotonous decrease of the Gilbert 
damping due to electron-phonon scattering with rising temperature. On the other hand, the curve representing the effect 
of the electron scattering due to thermal spin fluctuations exhibits a decrease in the low temperature region due to the 
dominating breathing Fermi surface dissipation mechanism, while above $~150$ K the increase of the Gilbert damping is 
determined by the increase of thermal magnetic disorder leading to magnetization dynamics due to electron scattering 
events accompanied by spin-flip electron transitions. However, approaching the Curie temperature, the Gilbert damping 
reaches a maximum at $~275$ K with a following decrease up to the Curie temperature. This behavior correlates with the 
temperature dependent behavior of the resistivity $\rho_{zz}(T)$and can be associated with the decrease of probability 
of spin-flip scattering of transport electrons caused by a modification of the electronic structure discussed 
above. A similar non-monotonous behavior has been found for the temperature dependence of the total Gilbert damping. 

\section{Conclusions \label{sec:C}}

In summary, we have studied the transport properties in the highly correlated system Gd from first principles. 
The electron-electron correlation effects were approximately accounted for by using the LSDA+$U$ approach resulting in 
an adequate description of the electronic structure. In turn, it enables a proper physical description of the 
transport properties. In this contribution we discussed the impact of finite temperatures (including the impact 
of thermal lattice vibrations and spin fluctuations) on the electrical resistivity as well as on the 
anomalous Hall resistivity. The applied approach based on the single site CPA describing thermal lattice vibrations and 
spin fluctuations allows to analyze individual contributions to the longitudinal and transverse resistivities arising 
due to these mechanisms. In both cases it turned out that in order to obtain reasonable agreement with experimental 
data it is necessary to account for a combination of the contributions connected with the phonon scattering and 
scattering by spin disorder as the simple sum of these contributions, especially for the AHR, significantly deviates 
from experiment. In the case of the longitudinal resistivity a slight anisotropy was observed which is in agreement with 
experimental results. For the out-of-plane resistivity a small experimentally detected maximum in the vicinity of the 
Curie temperature was fully reproduced. The emergence of this maximum according to experimental findings was 
attributed so far to magnetic short-range order effect. However, in the present calculations such an ordering was 
completely neglected as the distribution of the spin magnetic moments considered absolutely random. Accordingly, the 
origin of this maximum is solely due to spin disorder.

In case of the AHR a small anisotropy was observed as well. The calculated temperature dependent AHR with 
magnetization pointing along the $c$ axis agrees surprisingly well with the experimental data. The maximum 
occured just below the Curie temperature with the further abrupt drop is well reproduced.


\begin{acknowledgments}
This work was financially supported by the Deutsche Forschungsgemeinschaft (DFG) via SFB 689 and FOR1346 (DMFT). 
The authors would like to thank L. Szunyogh, L. Oroszl\'any and S. Chadov for fruitful discussions.
\end{acknowledgments}




\begin{thebibliography}{48}%
\makeatletter
\providecommand \@ifxundefined [1]{%
 \@ifx{#1\undefined}
}%
\providecommand \@ifnum [1]{%
 \ifnum #1\expandafter \@firstoftwo
 \else \expandafter \@secondoftwo
 \fi
}%
\providecommand \@ifx [1]{%
 \ifx #1\expandafter \@firstoftwo
 \else \expandafter \@secondoftwo
 \fi
}%
\providecommand \natexlab [1]{#1}%
\providecommand \enquote  [1]{``#1''}%
\providecommand \bibnamefont  [1]{#1}%
\providecommand \bibfnamefont [1]{#1}%
\providecommand \citenamefont [1]{#1}%
\providecommand \href@noop [0]{\@secondoftwo}%
\providecommand \href [0]{\begingroup \@sanitize@url \@href}%
\providecommand \@href[1]{\@@startlink{#1}\@@href}%
\providecommand \@@href[1]{\endgroup#1\@@endlink}%
\providecommand \@sanitize@url [0]{\catcode `\\12\catcode `\$12\catcode
  `\&12\catcode `\#12\catcode `\^12\catcode `\_12\catcode `\%12\relax}%
\providecommand \@@startlink[1]{}%
\providecommand \@@endlink[0]{}%
\providecommand \url  [0]{\begingroup\@sanitize@url \@url }%
\providecommand \@url [1]{\endgroup\@href {#1}{\urlprefix }}%
\providecommand \urlprefix  [0]{URL }%
\providecommand \Eprint [0]{\href }%
\providecommand \doibase [0]{http://dx.doi.org/}%
\providecommand \selectlanguage [0]{\@gobble}%
\providecommand \bibinfo  [0]{\@secondoftwo}%
\providecommand \bibfield  [0]{\@secondoftwo}%
\providecommand \translation [1]{[#1]}%
\providecommand \BibitemOpen [0]{}%
\providecommand \bibitemStop [0]{}%
\providecommand \bibitemNoStop [0]{.\EOS\space}%
\providecommand \EOS [0]{\spacefactor3000\relax}%
\providecommand \BibitemShut  [1]{\csname bibitem#1\endcsname}%
\let\auto@bib@innerbib\@empty
\bibitem [{\citenamefont {Belov}\ and\ \citenamefont {Ped'ko}(1962)}]{BP62}%
  \BibitemOpen
  \bibfield  {author} {\bibinfo {author} {\bibfnamefont {K.~P.}\ \bibnamefont
  {Belov}}\ and\ \bibinfo {author} {\bibfnamefont {A.~V.}\ \bibnamefont
  {Ped'ko}},\ }\href@noop {} {\bibfield  {journal} {\bibinfo  {journal} {Soviet
  Phys. JEPT}\ }\textbf {\bibinfo {volume} {15}},\ \bibinfo {pages} {62}
  (\bibinfo {year} {1962})}\BibitemShut {NoStop}%
\bibitem [{\citenamefont {Kn\"opfle}\ and\ \citenamefont
  {Sandratskii}(2000)}]{KS00}%
  \BibitemOpen
  \bibfield  {author} {\bibinfo {author} {\bibfnamefont {K.}~\bibnamefont
  {Kn\"opfle}}\ and\ \bibinfo {author} {\bibfnamefont {L.~M.}\ \bibnamefont
  {Sandratskii}},\ }\href {\doibase 10.1103/PhysRevB.63.014411} {\bibfield
  {journal} {\bibinfo  {journal} {Phys. Rev. B}\ }\textbf {\bibinfo {volume}
  {63}},\ \bibinfo {pages} {014411} (\bibinfo {year} {2000})}\BibitemShut
  {NoStop}%
\bibitem [{\citenamefont {Will}\ and\ \citenamefont {Alperin}(1964)}]{WNA64}%
  \BibitemOpen
  \bibfield  {author} {\bibinfo {author} {\bibfnamefont {N.~R.}\ \bibnamefont
  {Will}, \bibfnamefont {G.}}\ and\ \bibinfo {author} {\bibfnamefont {H.~A.}\
  \bibnamefont {Alperin}},\ }\href {http://dx.doi.org/10.1063/1.1713371}
  {\bibfield  {journal} {\bibinfo  {journal} {J. Appl. Physics}\ }\textbf
  {\bibinfo {volume} {35}},\ \bibinfo {pages} {1045} (\bibinfo {year}
  {1964})}\BibitemShut {NoStop}%
\bibitem [{\citenamefont {Trombe}(1937)}]{Trombe37}%
  \BibitemOpen
  \bibfield  {author} {\bibinfo {author} {\bibfnamefont {F.}~\bibnamefont
  {Trombe}},\ }\href@noop {} {\bibfield  {journal} {\bibinfo  {journal} {Ann.
  Phys.}\ }\textbf {\bibinfo {volume} {7}},\ \bibinfo {pages} {383} (\bibinfo
  {year} {1937})}\BibitemShut {NoStop}%
\bibitem [{\citenamefont {Elliott}\ and\ \citenamefont
  {Spedding}(1953)}]{ELS53}%
  \BibitemOpen
  \bibfield  {author} {\bibinfo {author} {\bibfnamefont {L.~S.}\ \bibnamefont
  {Elliott}, \bibfnamefont {J.~F.}}\ and\ \bibinfo {author} {\bibfnamefont
  {F.~H.}\ \bibnamefont {Spedding}},\ }\href
  {http://dx.doi.org/10.1103/PhysRev.91.28} {\bibfield  {journal} {\bibinfo
  {journal} {Phys. Rev.}\ }\textbf {\bibinfo {volume} {91}},\ \bibinfo {pages}
  {28} (\bibinfo {year} {1953})}\BibitemShut {NoStop}%
\bibitem [{\citenamefont {Nigh}\ \emph {et~al.}(1963)\citenamefont {Nigh},
  \citenamefont {Legvold},\ and\ \citenamefont {Spedding}}]{NLS63}%
  \BibitemOpen
  \bibfield  {author} {\bibinfo {author} {\bibfnamefont {H.~E.}\ \bibnamefont
  {Nigh}}, \bibinfo {author} {\bibfnamefont {S.}~\bibnamefont {Legvold}}, \
  and\ \bibinfo {author} {\bibfnamefont {F.~H.}\ \bibnamefont {Spedding}},\
  }\href {\doibase 10.1103/PhysRev.132.1092} {\bibfield  {journal} {\bibinfo
  {journal} {Phys. Rev.}\ }\textbf {\bibinfo {volume} {132}},\ \bibinfo {pages}
  {1092} (\bibinfo {year} {1963})}\BibitemShut {NoStop}%
\bibitem [{\citenamefont {Roeland}\ \emph {et~al.}(1975)\citenamefont
  {Roeland}, \citenamefont {Cock}, \citenamefont {Muller}, \citenamefont
  {Moleman}, \citenamefont {McEwen}, \citenamefont {Jordan},\ and\
  \citenamefont {Jones}}]{RCM+75}%
  \BibitemOpen
  \bibfield  {author} {\bibinfo {author} {\bibfnamefont {L.~W.}\ \bibnamefont
  {Roeland}}, \bibinfo {author} {\bibfnamefont {G.~J.}\ \bibnamefont {Cock}},
  \bibinfo {author} {\bibfnamefont {F.~A.}\ \bibnamefont {Muller}}, \bibinfo
  {author} {\bibfnamefont {A.~C.}\ \bibnamefont {Moleman}}, \bibinfo {author}
  {\bibfnamefont {K.~A.}\ \bibnamefont {McEwen}}, \bibinfo {author}
  {\bibfnamefont {R.~G.}\ \bibnamefont {Jordan}}, \ and\ \bibinfo {author}
  {\bibfnamefont {D.~W.}\ \bibnamefont {Jones}},\ }\href@noop {} {\bibfield
  {journal} {\bibinfo  {journal} {J. Phys. F}\ }\textbf {\bibinfo {volume}
  {5}},\ \bibinfo {pages} {L233} (\bibinfo {year} {1975})}\BibitemShut
  {NoStop}%
\bibitem [{\citenamefont {Bongsoo}\ \emph {et~al.}(1992)\citenamefont
  {Bongsoo}, \citenamefont {Andrews}, \citenamefont {Erskine}, \citenamefont
  {Kwang~Joo},\ and\ \citenamefont {Harmon}}]{BAE+92}%
  \BibitemOpen
  \bibfield  {author} {\bibinfo {author} {\bibfnamefont {K.}~\bibnamefont
  {Bongsoo}}, \bibinfo {author} {\bibfnamefont {A.~B.}\ \bibnamefont
  {Andrews}}, \bibinfo {author} {\bibfnamefont {J.~L.}\ \bibnamefont
  {Erskine}}, \bibinfo {author} {\bibfnamefont {K.}~\bibnamefont {Kwang~Joo}},
  \ and\ \bibinfo {author} {\bibfnamefont {B.~N.}\ \bibnamefont {Harmon}},\
  }\href {\doibase 10.1103/PhysRevLett.68.1931} {\bibfield  {journal} {\bibinfo
   {journal} {Phys. Rev. Lett.}\ }\textbf {\bibinfo {volume} {68}},\ \bibinfo
  {pages} {1931} (\bibinfo {year} {1992})}\BibitemShut {NoStop}%
\bibitem [{\citenamefont {Weschke}\ \emph {et~al.}(1996)\citenamefont
  {Weschke}, \citenamefont {Sch\"ussler-Langeheine}, \citenamefont {Meier},
  \citenamefont {Fedorov}, \citenamefont {Starke}, \citenamefont {H\"ubinger},\
  and\ \citenamefont {Kaindl}}]{WSM+96}%
  \BibitemOpen
  \bibfield  {author} {\bibinfo {author} {\bibfnamefont {E.}~\bibnamefont
  {Weschke}}, \bibinfo {author} {\bibfnamefont {C.}~\bibnamefont
  {Sch\"ussler-Langeheine}}, \bibinfo {author} {\bibfnamefont {R.}~\bibnamefont
  {Meier}}, \bibinfo {author} {\bibfnamefont {A.~V.}\ \bibnamefont {Fedorov}},
  \bibinfo {author} {\bibfnamefont {K.}~\bibnamefont {Starke}}, \bibinfo
  {author} {\bibfnamefont {F.}~\bibnamefont {H\"ubinger}}, \ and\ \bibinfo
  {author} {\bibfnamefont {G.}~\bibnamefont {Kaindl}},\ }\href {\doibase
  10.1103/PhysRevLett.77.3415} {\bibfield  {journal} {\bibinfo  {journal}
  {Phys. Rev. Lett.}\ }\textbf {\bibinfo {volume} {77}},\ \bibinfo {pages}
  {3415} (\bibinfo {year} {1996})}\BibitemShut {NoStop}%
\bibitem [{\citenamefont {Maiti}\ \emph {et~al.}(2002)\citenamefont {Maiti},
  \citenamefont {Malagoli}, \citenamefont {Dallmeyer},\ and\ \citenamefont
  {Carbone}}]{MMD+02}%
  \BibitemOpen
  \bibfield  {author} {\bibinfo {author} {\bibfnamefont {K.}~\bibnamefont
  {Maiti}}, \bibinfo {author} {\bibfnamefont {M.~C.}\ \bibnamefont {Malagoli}},
  \bibinfo {author} {\bibfnamefont {A.}~\bibnamefont {Dallmeyer}}, \ and\
  \bibinfo {author} {\bibfnamefont {C.}~\bibnamefont {Carbone}},\ }\href
  {\doibase 10.1103/PhysRevLett.88.167205} {\bibfield  {journal} {\bibinfo
  {journal} {Phys. Rev. Lett.}\ }\textbf {\bibinfo {volume} {88}},\ \bibinfo
  {pages} {167205} (\bibinfo {year} {2002})}\BibitemShut {NoStop}%
\bibitem [{\citenamefont {Glasbrenner}\ \emph {et~al.}(2012)\citenamefont
  {Glasbrenner}, \citenamefont {Belashchenko}, \citenamefont {Kudrnovsk\'y},
  \citenamefont {Drchal}, \citenamefont {Khmelevskyi},\ and\ \citenamefont
  {Turek}}]{GBK+12a}%
  \BibitemOpen
  \bibfield  {author} {\bibinfo {author} {\bibfnamefont {J.~K.}\ \bibnamefont
  {Glasbrenner}}, \bibinfo {author} {\bibfnamefont {K.~D.}\ \bibnamefont
  {Belashchenko}}, \bibinfo {author} {\bibfnamefont {J.}~\bibnamefont
  {Kudrnovsk\'y}}, \bibinfo {author} {\bibfnamefont {V.}~\bibnamefont
  {Drchal}}, \bibinfo {author} {\bibfnamefont {S.}~\bibnamefont {Khmelevskyi}},
  \ and\ \bibinfo {author} {\bibfnamefont {I.}~\bibnamefont {Turek}},\ }\href
  {\doibase 10.1103/PhysRevB.85.214405} {\bibfield  {journal} {\bibinfo
  {journal} {Phys. Rev. B}\ }\textbf {\bibinfo {volume} {85}},\ \bibinfo
  {pages} {214405} (\bibinfo {year} {2012})}\BibitemShut {NoStop}%
\bibitem [{\citenamefont {Oroszl\'any}\ \emph {et~al.}(2015)\citenamefont
  {Oroszl\'any}, \citenamefont {De\'ak}, \citenamefont {Simon}, \citenamefont
  {Khmelevskyi},\ and\ \citenamefont {Szunyogh}}]{ODS+15}%
  \BibitemOpen
  \bibfield  {author} {\bibinfo {author} {\bibfnamefont {L.}~\bibnamefont
  {Oroszl\'any}}, \bibinfo {author} {\bibfnamefont {A.}~\bibnamefont {De\'ak}},
  \bibinfo {author} {\bibfnamefont {E.}~\bibnamefont {Simon}}, \bibinfo
  {author} {\bibfnamefont {S.}~\bibnamefont {Khmelevskyi}}, \ and\ \bibinfo
  {author} {\bibfnamefont {L.}~\bibnamefont {Szunyogh}},\ }\href {\doibase
  10.1103/PhysRevLett.115.096402} {\bibfield  {journal} {\bibinfo  {journal}
  {Phys. Rev. Lett.}\ }\textbf {\bibinfo {volume} {115}},\ \bibinfo {pages}
  {096402} (\bibinfo {year} {2015})}\BibitemShut {NoStop}%
\bibitem [{\citenamefont {Khmelevskyi}\ \emph {et~al.}(2004)\citenamefont
  {Khmelevskyi}, \citenamefont {Turek},\ and\ \citenamefont {Mohn}}]{KTM04}%
  \BibitemOpen
  \bibfield  {author} {\bibinfo {author} {\bibfnamefont {S.}~\bibnamefont
  {Khmelevskyi}}, \bibinfo {author} {\bibfnamefont {I.}~\bibnamefont {Turek}},
  \ and\ \bibinfo {author} {\bibfnamefont {P.}~\bibnamefont {Mohn}},\ }\href
  {\doibase 10.1103/PhysRevB.70.132401} {\bibfield  {journal} {\bibinfo
  {journal} {Phys. Rev. B}\ }\textbf {\bibinfo {volume} {70}},\ \bibinfo
  {pages} {132401} (\bibinfo {year} {2004})}\BibitemShut {NoStop}%
\bibitem [{\citenamefont {Donath}\ \emph {et~al.}(1996)\citenamefont {Donath},
  \citenamefont {Gubanka},\ and\ \citenamefont {Passek}}]{DGP96}%
  \BibitemOpen
  \bibfield  {author} {\bibinfo {author} {\bibfnamefont {M.}~\bibnamefont
  {Donath}}, \bibinfo {author} {\bibfnamefont {B.}~\bibnamefont {Gubanka}}, \
  and\ \bibinfo {author} {\bibfnamefont {F.}~\bibnamefont {Passek}},\ }\href
  {\doibase 10.1103/PhysRevLett.77.5138} {\bibfield  {journal} {\bibinfo
  {journal} {Phys. Rev. Lett.}\ }\textbf {\bibinfo {volume} {77}},\ \bibinfo
  {pages} {5138} (\bibinfo {year} {1996})}\BibitemShut {NoStop}%
\bibitem [{\citenamefont {Li}\ \emph {et~al.}(1995)\citenamefont {Li},
  \citenamefont {Pearson}, \citenamefont {Bader}, \citenamefont {McIlroy},
  \citenamefont {Waldfried},\ and\ \citenamefont {Dowben}}]{LPB+95}%
  \BibitemOpen
  \bibfield  {author} {\bibinfo {author} {\bibfnamefont {D.}~\bibnamefont
  {Li}}, \bibinfo {author} {\bibfnamefont {J.}~\bibnamefont {Pearson}},
  \bibinfo {author} {\bibfnamefont {S.~D.}\ \bibnamefont {Bader}}, \bibinfo
  {author} {\bibfnamefont {D.~N.}\ \bibnamefont {McIlroy}}, \bibinfo {author}
  {\bibfnamefont {C.}~\bibnamefont {Waldfried}}, \ and\ \bibinfo {author}
  {\bibfnamefont {P.~A.}\ \bibnamefont {Dowben}},\ }\href {\doibase
  10.1103/PhysRevB.51.13895} {\bibfield  {journal} {\bibinfo  {journal} {Phys.
  Rev. B}\ }\textbf {\bibinfo {volume} {51}},\ \bibinfo {pages} {13895}
  (\bibinfo {year} {1995})}\BibitemShut {NoStop}%
\bibitem [{\citenamefont {Fedorov}\ \emph {et~al.}(2002)\citenamefont
  {Fedorov}, \citenamefont {Valla}, \citenamefont {Liu}, \citenamefont
  {Johnson}, \citenamefont {Weinert},\ and\ \citenamefont {Allen}}]{FVL+02}%
  \BibitemOpen
  \bibfield  {author} {\bibinfo {author} {\bibfnamefont {A.~V.}\ \bibnamefont
  {Fedorov}}, \bibinfo {author} {\bibfnamefont {T.}~\bibnamefont {Valla}},
  \bibinfo {author} {\bibfnamefont {F.}~\bibnamefont {Liu}}, \bibinfo {author}
  {\bibfnamefont {P.~D.}\ \bibnamefont {Johnson}}, \bibinfo {author}
  {\bibfnamefont {M.}~\bibnamefont {Weinert}}, \ and\ \bibinfo {author}
  {\bibfnamefont {P.~B.}\ \bibnamefont {Allen}},\ }\href {\doibase
  10.1103/PhysRevB.65.212409} {\bibfield  {journal} {\bibinfo  {journal} {Phys.
  Rev. B}\ }\textbf {\bibinfo {volume} {65}},\ \bibinfo {pages} {212409}
  (\bibinfo {year} {2002})}\BibitemShut {NoStop}%
\bibitem [{\citenamefont {Sandratskii}(2014)}]{San14}%
  \BibitemOpen
  \bibfield  {author} {\bibinfo {author} {\bibfnamefont {L.~M.}\ \bibnamefont
  {Sandratskii}},\ }\href {\doibase 10.1103/PhysRevB.90.184406} {\bibfield
  {journal} {\bibinfo  {journal} {Phys. Rev. B}\ }\textbf {\bibinfo {volume}
  {90}},\ \bibinfo {pages} {184406} (\bibinfo {year} {2014})}\BibitemShut
  {NoStop}%
\bibitem [{\citenamefont {Seib}\ and\ \citenamefont {F\"ahnle}(2010)}]{SF10}%
  \BibitemOpen
  \bibfield  {author} {\bibinfo {author} {\bibfnamefont {J.}~\bibnamefont
  {Seib}}\ and\ \bibinfo {author} {\bibfnamefont {M.}~\bibnamefont
  {F\"ahnle}},\ }\href {\doibase 10.1103/PhysRevB.82.064401} {\bibfield
  {journal} {\bibinfo  {journal} {Phys. Rev. B}\ }\textbf {\bibinfo {volume}
  {82}},\ \bibinfo {pages} {064401} (\bibinfo {year} {2010})}\BibitemShut
  {NoStop}%
\bibitem [{\citenamefont {Seib}\ \emph {et~al.}(2009)\citenamefont {Seib},
  \citenamefont {Steiauf},\ and\ \citenamefont {F\"ahnle}}]{SSF09}%
  \BibitemOpen
  \bibfield  {author} {\bibinfo {author} {\bibfnamefont {J.}~\bibnamefont
  {Seib}}, \bibinfo {author} {\bibfnamefont {D.}~\bibnamefont {Steiauf}}, \
  and\ \bibinfo {author} {\bibfnamefont {M.}~\bibnamefont {F\"ahnle}},\ }\href
  {\doibase 10.1103/PhysRevB.79.064419} {\bibfield  {journal} {\bibinfo
  {journal} {Phys. Rev. B}\ }\textbf {\bibinfo {volume} {79}},\ \bibinfo
  {pages} {064419} (\bibinfo {year} {2009})}\BibitemShut {NoStop}%
\bibitem [{\citenamefont {Savrasov}\ and\ \citenamefont
  {Savrasov}(1996)}]{SS96}%
  \BibitemOpen
  \bibfield  {author} {\bibinfo {author} {\bibfnamefont {S.~Y.}\ \bibnamefont
  {Savrasov}}\ and\ \bibinfo {author} {\bibfnamefont {D.~Y.}\ \bibnamefont
  {Savrasov}},\ }\href {\doibase 10.1103/PhysRevB.54.16487} {\bibfield
  {journal} {\bibinfo  {journal} {Phys. Rev. B}\ }\textbf {\bibinfo {volume}
  {54}},\ \bibinfo {pages} {16487} (\bibinfo {year} {1996})}\BibitemShut
  {NoStop}%
\bibitem [{\citenamefont {Kasuya}(1956)}]{Kas56}%
  \BibitemOpen
  \bibfield  {author} {\bibinfo {author} {\bibfnamefont {T.}~\bibnamefont
  {Kasuya}},\ }\href {\doibase 10.1143/PTP.16.58} {\bibfield  {journal}
  {\bibinfo  {journal} {Prog. Theor. Phys.}\ }\textbf {\bibinfo {volume}
  {16}},\ \bibinfo {pages} {58} (\bibinfo {year} {1956})}\BibitemShut {NoStop}%
\bibitem [{\citenamefont {Kasuya}(1959)}]{Kas59}%
  \BibitemOpen
  \bibfield  {author} {\bibinfo {author} {\bibfnamefont {T.}~\bibnamefont
  {Kasuya}},\ }\href {\doibase 10.1143/PTP.22.227} {\bibfield  {journal}
  {\bibinfo  {journal} {Prog. Theor. Phys.}\ }\textbf {\bibinfo {volume}
  {22}},\ \bibinfo {pages} {227} (\bibinfo {year} {1959})}\BibitemShut
  {NoStop}%
\bibitem [{\citenamefont {de~Gennes}\ and\ \citenamefont
  {Friedel}(1958)}]{GF58}%
  \BibitemOpen
  \bibfield  {author} {\bibinfo {author} {\bibfnamefont {P.}~\bibnamefont
  {de~Gennes}}\ and\ \bibinfo {author} {\bibfnamefont {J.}~\bibnamefont
  {Friedel}},\ }\href
  {http://www.sciencedirect.com/science/article/pii/0022369758901963}
  {\bibfield  {journal} {\bibinfo  {journal} {J. Phys. Chem. Solids}\ }\textbf
  {\bibinfo {volume} {4}},\ \bibinfo {pages} {71 } (\bibinfo {year}
  {1958})}\BibitemShut {NoStop}%
\bibitem [{\citenamefont {Kudrnovsk\'y}\ \emph {et~al.}(2012)\citenamefont
  {Kudrnovsk\'y}, \citenamefont {Drchal}, \citenamefont {Turek}, \citenamefont
  {Khmelevskyi}, \citenamefont {Glasbrenner},\ and\ \citenamefont
  {Belashchenko}}]{KDT+12}%
  \BibitemOpen
  \bibfield  {author} {\bibinfo {author} {\bibfnamefont {J.}~\bibnamefont
  {Kudrnovsk\'y}}, \bibinfo {author} {\bibfnamefont {V.}~\bibnamefont
  {Drchal}}, \bibinfo {author} {\bibfnamefont {I.}~\bibnamefont {Turek}},
  \bibinfo {author} {\bibfnamefont {S.}~\bibnamefont {Khmelevskyi}}, \bibinfo
  {author} {\bibfnamefont {J.~K.}\ \bibnamefont {Glasbrenner}}, \ and\ \bibinfo
  {author} {\bibfnamefont {K.~D.}\ \bibnamefont {Belashchenko}},\ }\href
  {\doibase 10.1103/PhysRevB.86.144423} {\bibfield  {journal} {\bibinfo
  {journal} {Phys. Rev. B}\ }\textbf {\bibinfo {volume} {86}},\ \bibinfo
  {pages} {144423} (\bibinfo {year} {2012})}\BibitemShut {NoStop}%
\bibitem [{\citenamefont {Christen}\ \emph {et~al.}(1979)\citenamefont
  {Christen}, \citenamefont {Giovannini},\ and\ \citenamefont
  {Sierro}}]{CGS79}%
  \BibitemOpen
  \bibfield  {author} {\bibinfo {author} {\bibfnamefont {M.}~\bibnamefont
  {Christen}}, \bibinfo {author} {\bibfnamefont {B.}~\bibnamefont
  {Giovannini}}, \ and\ \bibinfo {author} {\bibfnamefont {J.}~\bibnamefont
  {Sierro}},\ }\href {\doibase 10.1103/PhysRevB.20.4624} {\bibfield  {journal}
  {\bibinfo  {journal} {Phys. Rev. B}\ }\textbf {\bibinfo {volume} {20}},\
  \bibinfo {pages} {4624} (\bibinfo {year} {1979})}\BibitemShut {NoStop}%
\bibitem [{\citenamefont {Karplus}\ and\ \citenamefont
  {Luttinger}(1954)}]{KL54}%
  \BibitemOpen
  \bibfield  {author} {\bibinfo {author} {\bibfnamefont {R.}~\bibnamefont
  {Karplus}}\ and\ \bibinfo {author} {\bibfnamefont {J.~M.}\ \bibnamefont
  {Luttinger}},\ }\href {\doibase 10.1103/PhysRev.95.1154} {\bibfield
  {journal} {\bibinfo  {journal} {Phys. Rev.}\ }\textbf {\bibinfo {volume}
  {95}},\ \bibinfo {pages} {1154} (\bibinfo {year} {1954})}\BibitemShut
  {NoStop}%
\bibitem [{\citenamefont {Kondo}(1962)}]{Kon62}%
  \BibitemOpen
  \bibfield  {author} {\bibinfo {author} {\bibfnamefont {J.}~\bibnamefont
  {Kondo}},\ }\href {\doibase 10.1143/PTP.27.772} {\bibfield  {journal}
  {\bibinfo  {journal} {Prog. Theor. Phys.}\ }\textbf {\bibinfo {volume}
  {27}},\ \bibinfo {pages} {772} (\bibinfo {year} {1962})}\BibitemShut
  {NoStop}%
\bibitem [{\citenamefont {Maranzana}(1967)}]{Mar67}%
  \BibitemOpen
  \bibfield  {author} {\bibinfo {author} {\bibfnamefont {F.~E.}\ \bibnamefont
  {Maranzana}},\ }\href {\doibase 10.1103/PhysRev.160.421} {\bibfield
  {journal} {\bibinfo  {journal} {Phys. Rev.}\ }\textbf {\bibinfo {volume}
  {160}},\ \bibinfo {pages} {421} (\bibinfo {year} {1967})}\BibitemShut
  {NoStop}%
\bibitem [{\citenamefont {Berger}(1970)}]{Ber70}%
  \BibitemOpen
  \bibfield  {author} {\bibinfo {author} {\bibfnamefont {L.}~\bibnamefont
  {Berger}},\ }\href {\doibase 10.1103/PhysRevB.2.4559} {\bibfield  {journal}
  {\bibinfo  {journal} {Phys. Rev. B}\ }\textbf {\bibinfo {volume} {2}},\
  \bibinfo {pages} {4559} (\bibinfo {year} {1970})}\BibitemShut {NoStop}%
\bibitem [{\citenamefont {Berger}(1972)}]{Ber72}%
  \BibitemOpen
  \bibfield  {author} {\bibinfo {author} {\bibfnamefont {L.}~\bibnamefont
  {Berger}},\ }\href {\doibase 10.1103/PhysRevB.5.1862} {\bibfield  {journal}
  {\bibinfo  {journal} {Phys. Rev. B}\ }\textbf {\bibinfo {volume} {5}},\
  \bibinfo {pages} {1862} (\bibinfo {year} {1972})}\BibitemShut {NoStop}%
\bibitem [{\citenamefont {Asomoza}\ and\ \citenamefont {Reich}(1983)}]{AFR83}%
  \BibitemOpen
  \bibfield  {author} {\bibinfo {author} {\bibfnamefont {F.~A.}\ \bibnamefont
  {Asomoza}, \bibfnamefont {R.}}\ and\ \bibinfo {author} {\bibfnamefont
  {R.}~\bibnamefont {Reich}},\ }\href
  {http://www.sciencedirect.com/science/article/pii/0022508883900681}
  {\bibfield  {journal} {\bibinfo  {journal} {J. Less-Common Met.}\ }\textbf
  {\bibinfo {volume} {90}},\ \bibinfo {pages} {177} (\bibinfo {year}
  {1983})}\BibitemShut {NoStop}%
\bibitem [{\citenamefont {Ebert}\ \emph {et~al.}(2015)\citenamefont {Ebert},
  \citenamefont {Mankovsky}, \citenamefont {Chadova}, \citenamefont {Polesya},
  \citenamefont {Min\'{a}r},\ and\ \citenamefont {K\"odderitzsch}}]{EMC+15}%
  \BibitemOpen
  \bibfield  {author} {\bibinfo {author} {\bibfnamefont {H.}~\bibnamefont
  {Ebert}}, \bibinfo {author} {\bibfnamefont {S.}~\bibnamefont {Mankovsky}},
  \bibinfo {author} {\bibfnamefont {K.}~\bibnamefont {Chadova}}, \bibinfo
  {author} {\bibfnamefont {S.}~\bibnamefont {Polesya}}, \bibinfo {author}
  {\bibfnamefont {J.}~\bibnamefont {Min\'{a}r}}, \ and\ \bibinfo {author}
  {\bibfnamefont {D.}~\bibnamefont {K\"odderitzsch}},\ }\href {\doibase
  http://dx.doi.org/10.1103/PhysRevB.91.165132} {\bibfield  {journal} {\bibinfo
   {journal} {Phys. Rev. B}\ }\textbf {\bibinfo {volume} {91}},\ \bibinfo
  {pages} {165132} (\bibinfo {year} {2015})}\BibitemShut {NoStop}%
\bibitem [{\citenamefont {Ebert}\ \emph
  {et~al.}(2011{\natexlab{a}})\citenamefont {Ebert}, \citenamefont
  {K\"odderitzsch},\ and\ \citenamefont {Min\'{a}r}}]{EKM11}%
  \BibitemOpen
  \bibfield  {author} {\bibinfo {author} {\bibfnamefont {H.}~\bibnamefont
  {Ebert}}, \bibinfo {author} {\bibfnamefont {D.}~\bibnamefont
  {K\"odderitzsch}}, \ and\ \bibinfo {author} {\bibfnamefont {J.}~\bibnamefont
  {Min\'{a}r}},\ }\href {\doibase 10.1088/0034-4885/74/9/096501} {\bibfield
  {journal} {\bibinfo  {journal} {Rep. Prog. Phys.}\ }\textbf {\bibinfo
  {volume} {74}},\ \bibinfo {pages} {096501} (\bibinfo {year}
  {2011}{\natexlab{a}})}\BibitemShut {NoStop}%
\bibitem [{\citenamefont {Engel}\ and\ \citenamefont {Dreizler}(2011)}]{ED11}%
  \BibitemOpen
  \bibfield  {author} {\bibinfo {author} {\bibfnamefont {E.}~\bibnamefont
  {Engel}}\ and\ \bibinfo {author} {\bibfnamefont {R.~M.}\ \bibnamefont
  {Dreizler}},\ }\href {\doibase 10.1007/978-3-642-14090-7} {\emph {\bibinfo
  {title} {Density Functional Theory -- An advanced course}}}\ (\bibinfo
  {publisher} {Springer},\ \bibinfo {address} {Berlin},\ \bibinfo {year}
  {2011})\BibitemShut {NoStop}%
\bibitem [{\citenamefont {Vosko}\ \emph {et~al.}(1980)\citenamefont {Vosko},
  \citenamefont {Wilk},\ and\ \citenamefont {Nusair}}]{VWN80}%
  \BibitemOpen
  \bibfield  {author} {\bibinfo {author} {\bibfnamefont {S.~H.}\ \bibnamefont
  {Vosko}}, \bibinfo {author} {\bibfnamefont {L.}~\bibnamefont {Wilk}}, \ and\
  \bibinfo {author} {\bibfnamefont {M.}~\bibnamefont {Nusair}},\ }\href
  {\doibase 10.1139/p80-159} {\bibfield  {journal} {\bibinfo  {journal} {Can.
  J. Phys.}\ }\textbf {\bibinfo {volume} {58}},\ \bibinfo {pages} {1200}
  (\bibinfo {year} {1980})}
  \BibitemShut
  {NoStop}%
\bibitem [{\citenamefont {Czy\.{z}yk}\ and\ \citenamefont
  {Sawatzky}(1994)}]{CS94}%
  \BibitemOpen
  \bibfield  {author} {\bibinfo {author} {\bibfnamefont {M.~T.}\ \bibnamefont
  {Czy\.{z}yk}}\ and\ \bibinfo {author} {\bibfnamefont {G.~A.}\ \bibnamefont
  {Sawatzky}},\ }\href {\doibase 10.1103/PhysRevB.49.14211} {\bibfield
  {journal} {\bibinfo  {journal} {Phys. Rev. B}\ }\textbf {\bibinfo {volume}
  {49}},\ \bibinfo {pages} {14211} (\bibinfo {year} {1994})}\BibitemShut
  {NoStop}%
\bibitem [{\citenamefont {Velick\'y}(1969)}]{Vel69}%
  \BibitemOpen
  \bibfield  {author} {\bibinfo {author} {\bibfnamefont {B.}~\bibnamefont
  {Velick\'y}},\ }\href {\doibase 10.1103/PhysRev.184.614} {\bibfield
  {journal} {\bibinfo  {journal} {Phys. Rev.}\ }\textbf {\bibinfo {volume}
  {184}},\ \bibinfo {pages} {614} (\bibinfo {year} {1969})}\BibitemShut
  {NoStop}%
\bibitem [{\citenamefont {Butler}(1985)}]{But85}%
  \BibitemOpen
  \bibfield  {author} {\bibinfo {author} {\bibfnamefont {W.~H.}\ \bibnamefont
  {Butler}},\ }\href {\doibase 10.1103/PhysRevB.31.3260} {\bibfield  {journal}
  {\bibinfo  {journal} {Phys. Rev. B}\ }\textbf {\bibinfo {volume} {31}},\
  \bibinfo {pages} {3260} (\bibinfo {year} {1985})}\BibitemShut {NoStop}%
\bibitem [{\citenamefont {Turek}\ \emph {et~al.}(2002)\citenamefont {Turek},
  \citenamefont {Kudrnovsk\'y}, \citenamefont {Drchal}, \citenamefont
  {Szunyogh},\ and\ \citenamefont {Weinberger}}]{TKD+02b}%
  \BibitemOpen
  \bibfield  {author} {\bibinfo {author} {\bibfnamefont {I.}~\bibnamefont
  {Turek}}, \bibinfo {author} {\bibfnamefont {J.}~\bibnamefont {Kudrnovsk\'y}},
  \bibinfo {author} {\bibfnamefont {V.}~\bibnamefont {Drchal}}, \bibinfo
  {author} {\bibfnamefont {L.}~\bibnamefont {Szunyogh}}, \ and\ \bibinfo
  {author} {\bibfnamefont {P.}~\bibnamefont {Weinberger}},\ }\href {\doibase
  10.1103/PhysRevB.65.125101} {\bibfield  {journal} {\bibinfo  {journal} {Phys.
  Rev. B}\ }\textbf {\bibinfo {volume} {65}},\ \bibinfo {pages} {125101}
  (\bibinfo {year} {2002})}\BibitemShut {NoStop}%
\bibitem [{\citenamefont {Naito}\ \emph {et~al.}(2010)\citenamefont {Naito},
  \citenamefont {Hirashima},\ and\ \citenamefont {Kontani}}]{NHK10}%
  \BibitemOpen
  \bibfield  {author} {\bibinfo {author} {\bibfnamefont {T.}~\bibnamefont
  {Naito}}, \bibinfo {author} {\bibfnamefont {D.~S.}\ \bibnamefont
  {Hirashima}}, \ and\ \bibinfo {author} {\bibfnamefont {H.}~\bibnamefont
  {Kontani}},\ }\href {\doibase 10.1103/PhysRevB.81.195111} {\bibfield
  {journal} {\bibinfo  {journal} {Phys. Rev. B}\ }\textbf {\bibinfo {volume}
  {81}},\ \bibinfo {pages} {195111} (\bibinfo {year} {2010})}\BibitemShut
  {NoStop}%
\bibitem [{\citenamefont {Brataas}\ \emph {et~al.}(2008)\citenamefont
  {Brataas}, \citenamefont {Tserkovnyak},\ and\ \citenamefont {Bauer}}]{BTB08}%
  \BibitemOpen
  \bibfield  {author} {\bibinfo {author} {\bibfnamefont {A.}~\bibnamefont
  {Brataas}}, \bibinfo {author} {\bibfnamefont {Y.}~\bibnamefont
  {Tserkovnyak}}, \ and\ \bibinfo {author} {\bibfnamefont {G.~E.~W.}\
  \bibnamefont {Bauer}},\ }\href {\doibase 10.1103/PhysRevLett.101.037207}
  {\bibfield  {journal} {\bibinfo  {journal} {Phys. Rev. Lett.}\ }\textbf
  {\bibinfo {volume} {101}},\ \bibinfo {pages} {037207} (\bibinfo {year}
  {2008})}\BibitemShut {NoStop}%
\bibitem [{\citenamefont {Ebert}\ \emph
  {et~al.}(2011{\natexlab{b}})\citenamefont {Ebert}, \citenamefont {Mankovsky},
  \citenamefont {K\"odderitzsch},\ and\ \citenamefont {Kelly}}]{EMKK11}%
  \BibitemOpen
  \bibfield  {author} {\bibinfo {author} {\bibfnamefont {H.}~\bibnamefont
  {Ebert}}, \bibinfo {author} {\bibfnamefont {S.}~\bibnamefont {Mankovsky}},
  \bibinfo {author} {\bibfnamefont {D.}~\bibnamefont {K\"odderitzsch}}, \ and\
  \bibinfo {author} {\bibfnamefont {P.~J.}\ \bibnamefont {Kelly}},\ }\href
  {\doibase 10.1103/PhysRevLett.107.066603} {\bibfield  {journal} {\bibinfo
  {journal} {Phys. Rev. Lett.}\ }\textbf {\bibinfo {volume} {107}},\ \bibinfo
  {pages} {066603} (\bibinfo {year} {2011}{\natexlab{b}})}\BibitemShut
  {NoStop}%
\bibitem [{\citenamefont {Lang}\ \emph {et~al.}(1981)\citenamefont {Lang},
  \citenamefont {Baer},\ and\ \citenamefont {Cox}}]{LBC81}%
  \BibitemOpen
  \bibfield  {author} {\bibinfo {author} {\bibfnamefont {J.~K.}\ \bibnamefont
  {Lang}}, \bibinfo {author} {\bibfnamefont {Y.}~\bibnamefont {Baer}}, \ and\
  \bibinfo {author} {\bibfnamefont {P.~A.}\ \bibnamefont {Cox}},\ }\href
  {\doibase 10.1088/0305-4608/11/1/015} {\bibfield  {journal} {\bibinfo
  {journal} {J. Phys. F: Met. Phys.}\ }\textbf {\bibinfo {volume} {11}},\
  \bibinfo {pages} {121} (\bibinfo {year} {1981})}\BibitemShut {NoStop}%
\bibitem [{\citenamefont {Kim}\ \emph {et~al.}(1992)\citenamefont {Kim},
  \citenamefont {Andrews}, \citenamefont {Erskine}, \citenamefont {Kim},\ and\
  \citenamefont {Harmon}}]{KAE+92}%
  \BibitemOpen
  \bibfield  {author} {\bibinfo {author} {\bibfnamefont {B.}~\bibnamefont
  {Kim}}, \bibinfo {author} {\bibfnamefont {A.~B.}\ \bibnamefont {Andrews}},
  \bibinfo {author} {\bibfnamefont {J.~L.}\ \bibnamefont {Erskine}}, \bibinfo
  {author} {\bibfnamefont {K.~J.}\ \bibnamefont {Kim}}, \ and\ \bibinfo
  {author} {\bibfnamefont {B.~N.}\ \bibnamefont {Harmon}},\ }\href {\doibase
  10.1103/PhysRevLett.68.1931} {\bibfield  {journal} {\bibinfo  {journal}
  {Phys. Rev. Lett.}\ }\textbf {\bibinfo {volume} {68}},\ \bibinfo {pages}
  {1931} (\bibinfo {year} {1992})}\BibitemShut {NoStop}%
\bibitem [{\citenamefont {Li}\ \emph {et~al.}(1992)\citenamefont {Li},
  \citenamefont {Zhang}, \citenamefont {Dowben},\ and\ \citenamefont
  {Onellion}}]{LZD+92}%
  \BibitemOpen
  \bibfield  {author} {\bibinfo {author} {\bibfnamefont {D.}~\bibnamefont
  {Li}}, \bibinfo {author} {\bibfnamefont {J.}~\bibnamefont {Zhang}}, \bibinfo
  {author} {\bibfnamefont {P.~A.}\ \bibnamefont {Dowben}}, \ and\ \bibinfo
  {author} {\bibfnamefont {M.}~\bibnamefont {Onellion}},\ }\href
  {http://link.aps.org/doi/10.1103/PhysRevB.45.7272} {\bibfield  {journal}
  {\bibinfo  {journal} {Phys. Rev. B}\ }\textbf {\bibinfo {volume} {45}},\
  \bibinfo {pages} {7272} (\bibinfo {year} {1992})}\BibitemShut {NoStop}%
\bibitem [{\citenamefont {Babushkina}(1966)}]{Bab66}%
  \BibitemOpen
  \bibfield  {author} {\bibinfo {author} {\bibfnamefont {N.}~\bibnamefont
  {Babushkina}},\ }\href@noop {} {\bibfield  {journal} {\bibinfo  {journal}
  {Sov. Phys. - Solid State}\ }\textbf {\bibinfo {volume} {7}},\ \bibinfo
  {pages} {2450} (\bibinfo {year} {1966})}\BibitemShut {NoStop}%
\bibitem [{\citenamefont {Lee}\ and\ \citenamefont {Legvold}(1967)}]{LL67}%
  \BibitemOpen
  \bibfield  {author} {\bibinfo {author} {\bibfnamefont {R.~S.}\ \bibnamefont
  {Lee}}\ and\ \bibinfo {author} {\bibfnamefont {S.}~\bibnamefont {Legvold}},\
  }\href {\doibase 10.1103/PhysRev.162.431} {\bibfield  {journal} {\bibinfo
  {journal} {Phys. Rev.}\ }\textbf {\bibinfo {volume} {162}},\ \bibinfo {pages}
  {431} (\bibinfo {year} {1967})}\BibitemShut {NoStop}%
\bibitem [{\citenamefont {N.~V.~Volkenshtein}\ and\ \citenamefont
  {Fedorov}(1966)}]{VGF70}%
  \BibitemOpen
  \bibfield  {author} {\bibinfo {author} {\bibfnamefont {I.~K.~G.}\
  \bibnamefont {N.~V.~Volkenshtein}}\ and\ \bibinfo {author} {\bibfnamefont
  {G.~V.}\ \bibnamefont {Fedorov}},\ }\href@noop {} {\bibfield  {journal}
  {\bibinfo  {journal} {JEPT}\ }\textbf {\bibinfo {volume} {23}},\ \bibinfo
  {pages} {1003} (\bibinfo {year} {1966})}\BibitemShut {NoStop}%
\end{thebibliography}
\end{document}